\documentclass{aastex}
\usepackage{emulateapj5}

\shorttitle{Spectral variability of PKS~0558--504}
\shortauthors{Gliozzi et al.}

  \def\pks{PKS~0558--504}

  \def\xmm{{\it XMM-Newton}} 
  \def\asca{{\it ASCA}} 
   
  \def\rxte{{\it RXTE}}

  \def\rosat{{\it ROSAT}}

  \def\lum{erg s$^{-1}$}

  \def\ltsima{$\; \buildrel < \over \sim \;$}
  \def\simlt{\lower.5ex\hbox{\ltsima}} 
  \def\gtsima{$\; \buildrel > \over \sim \;$}
  \def\simgt{\lower.5ex\hbox{\gtsima}} 

\begin{document}
\title{Long-Term X-Ray spectral variability of the radio-loud NLS1 galaxy PKS~0558--504}

\author{M. Gliozzi}
\affil{George Mason University, 4400 University Drive, Fairfax, VA 22030}

\author{I.E. Papadakis}
\affil{Physics Department, University of Crete, 710 03 Heraklion,
Crete, Greece}

\author{W.P. Brinkmann}
\affil{Max-Planck-Institut f\"ur 
extraterrestrische Physik, Postfach 1312, D-85741 Garching, Germany}

\begin{abstract}
We present X-ray observations of the Narrow-Line Seyfert 1 
galaxy PKS~0558--504 obtained with \rxte\ during a 1-year monitoring
campaign. This source, which is one of the very few radio-loud NLS1 
galaxies, shows strong X-ray flux variability on time scales of weeks-months
accompanied by spectral variability. The main goal of this study is 
to investigate the spectral variability with model-independent methods
and time-resolved spectroscopy in order to shed light on the origin of
the X-rays. The main results can be summarized as follows: 1) The flux
typically changes by a factor of 1.5-2 on time scales of 10-30 days, with few
extreme events where the flux increases by a factor of $\sim$4 in 3 days.
2) We do not observe any large amplitude, flux related spectral variations. 
During the flux variations, the spectrum varies mainly in
normalization and not in shape. We do observe some small amplitude
spectral variations, which do not correlate with flux, although there is
a hint of spectral hardening as the source brightens.  
3) There is no evidence for reprocessing features such as the Fe K$\alpha$ line
or a Compton hump. We argue that \pks\ is a peculiar
object that appears to be different from most of the radio-quiet and radio-loud AGN.
If a jet is responsible for the bulk of the X-rays, it must operate in an unusual 
way. If instead a corona is responsible for the X-rays, the system might be a
large-scale analog of the Galactic black holes in
the transient intermediate state.
\end{abstract}


\section{Introduction}
Narrow-Line Seyfert 1 galaxies (NLS1) are identified by their optical emission 
line properties: the ratio [O III]/H$\beta$ is less than 3 and FWHM H$\beta$ is
less than 2000${~\rm km~s^{-1}}$ (Osterbrock \& Pogge 1985, Goodrich 1989). 
NLS1 exhibit characteristic features at other wavelengths as well: they are usually 
strong infrared emitters (Moran et al. 1996) and they are seldom 
radio loud (Ulvestad et al. 1995, Siebert et al. 1999, Grupe et al. 1999,
2000). The latter result has been recently confirmed by Komossa et al. (2006)
who performed a systematic search of radio-loud NLS1s, finding that less than
7\% have radio-loudness factors $R > 10$ and only 2.5\% have $R > 100$.
In the X-rays NLS1 have been generally found to have extreme spectral and variability
properties (e.g., Puchnarewicz et al. 1992; Forster \& Halpern 1996; Boller et al. 1997; 
Brandt et al. 1999; Leighly 1999a,b; Grupe et al. 2001).
It is now generally accepted that NLS1 are AGN in their early phase (Grupe et al. 1999; 
Mathur 2000), characterized by relatively small black hole masses (e.g.,  Grupe \& Mathur 2004),
and very high accretion rates in terms of Eddington units (e.g., Boroson \& Green 1992;
Sulentic et al. 2000; Grupe 2004).

PKS~0558--504 ($z=0.137, m_{\rm B}=14.97$) is one of the few radio-loud 
NLS1 galaxies ($R_{\rm L}=f_{\rm 5 GHz}/f_{\rm B}\simeq 27$, Siebert et al.
1999). It was optically identified on the basis of X-ray positions
from the High Energy Astronomy Observatory (HEAO-1, Remillard et al. 1986).
Based on the empirical correlation between the broad line region (BLR) and the optical
luminosity (Kaspi et al. 2000), and assuming that the BLR is virialized, Wang
et al. (2001) derived a black hole mass of $4.5\times 10^7~M_\odot$ for PKS~0558--504.

Previous X-ray observations with different satellites have confirmed 
that \pks\ shows the characteristic NLS1 properties: strong variability,
steep X-ray spectrum, substantial soft excess,and relatively high luminosity. 
By comparing the X-ray 
observations throughout the last decade, it 
is evident that the strong X-ray variability of PKS~0558--504 occurs 
persistently but without being accompanied by significant spectral
variability: the spectrum above 2 keV is consistently described by a
power-law model with photon index  $\Gamma\sim2.2$ (Gliozzi et al. 2000). 

PKS~0558--504 has been observed repeatedly by \xmm\ as a calibration and
performance verification target. A spectral analysis of the first \xmm\
observations (O'Brien et al. 2001) indicates that the broad-band spectrum is well fitted 
by a Comptonization model.  This is in substantial agreement with the findings of
Brinkmann et al. (2004), who, analyzing 
several \xmm\ observations of \pks, proposed a scenario with two Comptonization 
components: one with high temperature and low optical depth ($kT$ \gtsima 50 keV,
$\tau\sim 0.7$), typical of ``normal" broad-line Seyfert galaxies, and an
additional cooler component with larger optical depth ($kT \sim 4.5$ keV,
$\tau$ \gtsima 2). On the other hand, this interpretation is in sharp contrast
with the scenario proposed by Ballantyne et al. (2001) and Crummy et al. (2006) 
who hypothesize the dominant role played by a highly ionized reflector,
based on ASCA and \xmm\ spectral data, respectively. 
In any case, since the two competing models have spectral parameters 
poorly constrained, neither of them can be firmly ruled out.

A temporal analysis of the first \xmm\ observations of PKS~0558--504
(Gliozzi et al. 2001) showed persistent X-ray variability with a tendency of the
X-ray spectrum to harden when the count rate increases. A similar spectral
variability has been observed in another radio-loud NLS1, RX J0134-4258
(Grupe et al. 2000) and in the NLS1 1H 0707-495 (Gallo et al. 2004).
This trend is rather
unusual for Seyfert galaxies, which generally show an opposite behavior (e.g.,
Papadakis et al. 2002), instead, it is more typical
for jet-dominated sources (e.g., Fossati et al. 2000; Gliozzi et al. 2006a). 
Indeed, the contribution of the beamed emission from a putative jet has been 
proposed in the past (Remillard et al. 1991) as possible cause for the brightness and 
extreme short-term variability 
of PKS~0558--504 in the X-rays. However, Brinkmann et al. (2004) have shown that this
unusual spectral trend  seem to be present only sporadically and preferentially at 
relatively high count rates. This suggests
that the X-ray emission from PKS~0558--504 can be the results of a
``normal" Seyfert-like and a jet component emerging only during the high state.

In this paper, we start a systematic study of the long-term X-ray flux 
and spectral variability of \pks\ using proprietary Rossi X-ray
Timing Explorer (\rxte) observations in the 2-15 keV range (i.e., in an
energy band beyond the X-ray soft excess).
We use several
model-independent methods to study the X-ray temporal and 
spectral properties of this source. 
One of the main purposes of this analysis is to shed light on the origin
of the X-rays and in particular on the role
played by a jet in the X-rays.  Once the jet contribution is properly 
assessed, the physical parameters characterizing the accretion process onto the
supermassive black hole can be better constrained, and hence it is
possible to discriminate between competing theoretical models. 

This
model-independent characterization of a possible jet contribution at X-rays
is important not only for radio-loud AGN, but also for their
scaled-down counterparts, the Galactic black holes (GBHs), which  are
known to have compact radio jets during their ``low-hard" spectral state.
Also in the ``very-high" state, radio emission is observed
but in the form of optically thin radio flares 
which are thought to be signatures of powerful ejection events often 
resolved as extended jets (e.g., Mirabel \& Rodriguez 1994).
Despite the much higher signal-to-noise of
their spectra and light curves, the jet contribution to the total
X-ray emission in GBHs is still matter of 
strong debate (see, e.g., Markoff et al. 2003;
Zdziarski et al. 2004).

The outline of the paper is as follows. In $\S~2$ we describe the
observations and data reduction. The main characteristics of the 
X-ray light curve are described in $\S~3$. In $\S~4$ we study the X-ray
spectral variability of \pks\ with several model-independent methods. 
In $\S~5$ we describe the results of a time-resolved spectral analysis. 
In $\S~6$ we summarize the main results and discuss their implications.

\section{Observations and Data Reduction}

We use proprietary \rxte\ data of \pks\ that was observed between 2005 March 4
and 2006 March 2. A second monitoring campaign is currently underway. Both
campaigns are performed with similar sampling: for 11 months  \pks\ is
regularly observed for $\sim$ 1000--2000 s once every 2 days, while for one
month 2 pointings per day are performed.  The observations  were carried out
with the Proportional Counter Array (PCA; Jahoda et al. 1996),  and the
High-Energy X-Ray Timing Experiment (HEXTE; Rotschild et al. 1998) on \rxte.
Here we will consider only PCA data, because the signal-to-noise of the HEXTE
data is too low for a meaningful analysis.

The PCA  data were screened according to the following acceptance criteria: the
satellite was out of the South Atlantic Anomaly (SAA) for at least 30 minutes,
the Earth elevation angle was $\geq 10^{\circ}$, the offset from the nominal
optical position was $\leq 0^{\circ}\!\!.02$, and the parameter ELECTRON-2 was
$\leq 0.1$. The last criterion removes data with high particle background rates
in the Proportional Counter Units (PCUs). The PCA background spectra and light
curves were determined using the ${\rm L}7-240$ model developed at the \rxte\
Guest Observer Facility (GOF) and implemented by the program {\tt pcabackest}
v.2.1b.   This model is appropriate for ``faint'' sources, i.e., those
producing count rates less than 40 ${\rm s^{-1}~PCU^{-1}}$. 

All the above tasks were carried out using the {\tt FTOOLS} v.5.3.1 software
package and with the help of the  \verb+REX+  script provided by the \rxte\
GOF. Data were initially extracted with 16 s time resolution and subsequently
re-binned at different bin widths depending on the application.  The current
temporal analysis is restricted to PCA, STANDARD-2 mode, 2--15 keV, Layer 1
data, because that is where the PCA is best calibrated and most sensitive. PCUs
0 and 2 were turned on throughout the monitoring campaign. However, since the
propane layer on PCU0 was damaged in May 2000, causing  a systematic increase
of the background, we conservatively use only PCU2 for our analysis (see
below). All quoted count rates are therefore for one PCU.

The spectral analysis of PCA data was performed using the {\tt XSPEC v.11.3.1}
software package (Arnaud 1996). We used PCA response matrices and effective
area curves created specifically for the individual observations by the program
{\tt pcarsp}, taking into account the evolution of the detector properties. 
All the spectra were re-binned so that each bin contained at least 20 counts for the
$\chi^2$ statistic to be valid. Fits were performed in the energy range 3--15
keV, where the signal-to-noise ratio is the highest.

\section{The X-ray Light Curves}

\begin{figure}
\begin{center}
\includegraphics[bb=80 30 400 600,clip=,angle=0,width=12cm]{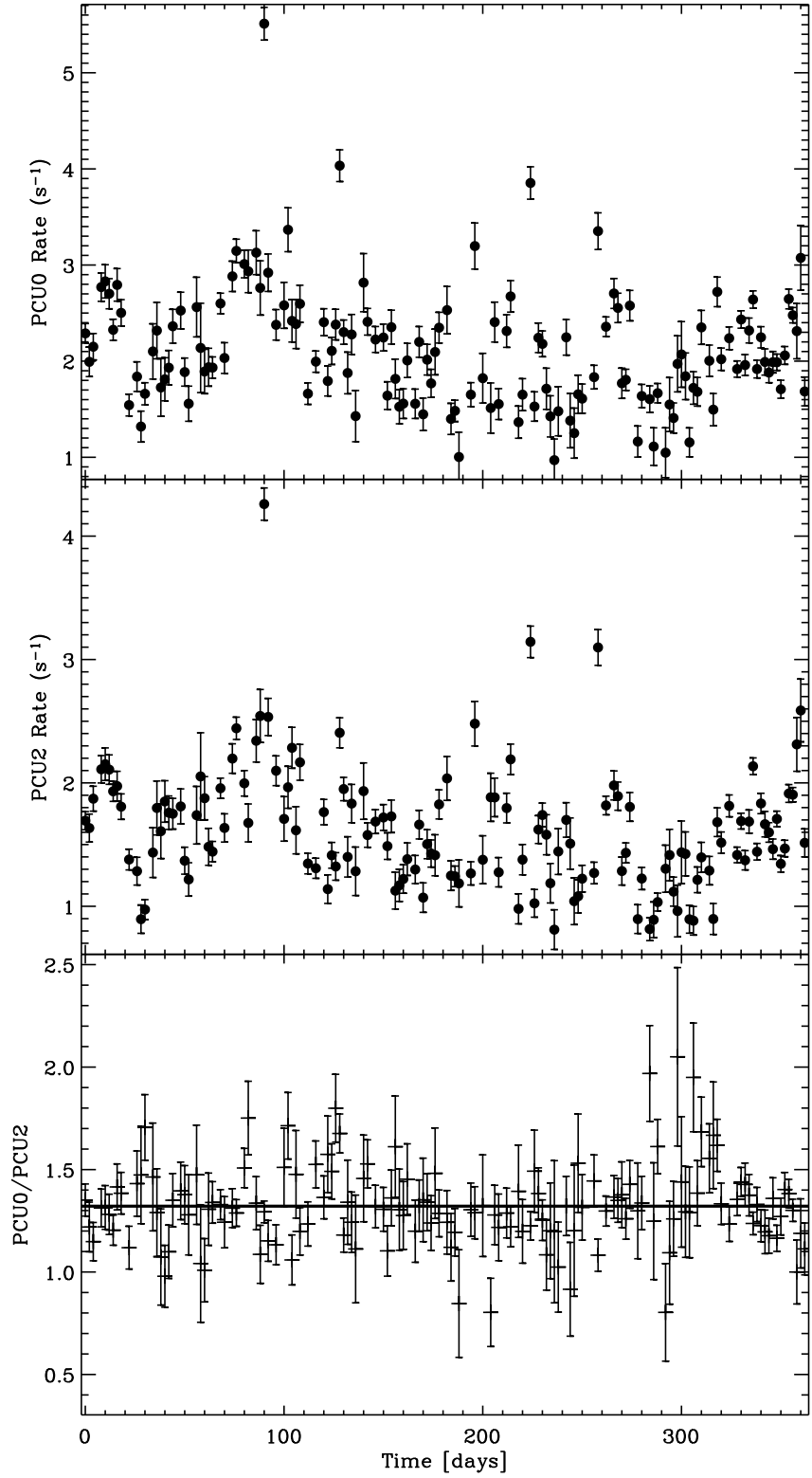}
\end{center}
\caption{Top panel: PCU0 light curve of \pks\ in the 2--15 keV range.
Middle panel: PCU2 light curve of \pks\ in the 2--15 keV range.
Bottom panel: PCU0/PCU2 light curve; the solid line represents the average value
of the ratio PCU0/PCU2. Time bins are two days. The reference time corresponding to
Time = 0 is 2005 March 4 UT 09:05:15.}
\label{figure:fig1}
\end{figure}

Figure~\ref{figure:fig1} shows the background-subtracted, 2--15 keV  light
curves of PCU0 (top panel), PCU2 (middle panel), and their ratio (bottom panel)
with a time bin of 2 days. A visual inspection of the bottom panel in this
Figure  suggests that the PCU0 and PCU2 light curves are broadly consistent
with each other, with the former being characterized by a higher count rate.
However, a formal check based on a $\chi^2$ test  indicates that the ratio
PCU0/PCU2 is not consistent  with the hypothesis of constancy  ($\chi^2=222.7$
for 143 degrees of freedom, hereafter dof). Since the propane layer on PCU0 was
damaged a few years ago, we decided to work with the PCU2 data only. We note
though that we also performed the same data analysis on the PCU0 data, and the
results are in broad agreement with those from the PCU2. 

Both the PCU0 and PCU2 light curves look similar to other long-term, RXTE
monitoring light curves of Seyfert galaxies (see Markowitz \& Edelson 2001 for
a review).   There are variations on all sampled time scales, with typical
amplitude (max-min)/min of the order of  $\sim$1.5--2 over 1 week -- 10 days,
and several peaks with the most prominent occurring $\sim$ 100 days after the
start of the monitoring campaign. These variations appear to be  erratic, and
we cannot easily identify any well defined, long-term trend, like
``flare-like" events, or long-term flux decay/rise trends. The analysis of the
flux variations using ``traditional" (i.e. power spectrum analysis) and
``non-traditional" techniques (e.g., Gliozzi et al. 2006b) will be presented in
the future after completion  of the second monitoring campaign. Here we limit
ourselves to the analysis of the spectral variability of the first year of
\rxte\ observations of \pks.

To this end, we constructed background subtracted light curves in two energy
bands, namely the ``soft" (2.5--5 keV) and the ``hard" (5--15 keV) band.  We
show them in Figure~\ref{figure:fig2}, together with the  hardness ratio
(hard/soft). Time bins are two days. As mentioned in
the previous section, the \pks\ \rxte\ observations were carried out following
two different sampling patterns, in order to probe as many time scales as
possible and hence estimate the power spectral density as accurately as
possible.
It is likely that  in disk and/or jet systems, different physical processes
may be at work on different time scales. For example, one may expect that the
quite different (in magnitude) viscous, thermal, and dynamical time scales
(associated with the accretion process),  heating and reprocessing time scales
(associated with the disk-corona interactions), or, in the case of
jet-dominated objects, the  cooling and acceleration time scales of relativistic
electrons, may play a role. For this reason, we first treat separately the 
data from ``sampling period 1" (hereafter S1: eleven months with one observation
every 2 days) and ``sampling period 2" (S2: one month with two observations per day),
using time bins of 5760 s ($\sim$ 1 \rxte\ orbit). If the results are consistent
(as it is the case for \pks; see below), we combine the two data sets and repeat the
analysis using time bins of two days to increase the significance of the results.
 
The soft and hard band light curves  appear to be quite similar, with the same
variations appearing in them.
This is also implied by the hardness ratio light curve, which appears to be roughly
constant.  However, a $\chi^2$ test against the hypothesis of constant hardness
ratio implies significant spectral variations. We find that $\chi^2=224.4/140$ dof
(using separate samplings: $\chi^2=247.3/166$ dof  and  $\chi^2=116/53$ dof 
for the S1 and S2 data, respectively). 
We have computed the fractional variability (see $\S4.3$ for more details) for
the soft, hard, and $HR$ light curves:
$F_{\rm var,soft}=0.31\pm0.01$, $F_{\rm var,hard}=0.29\pm0.01$,
$F_{\rm var,HR}=0.08\pm0.02$. 
The variability amplitude is roughly similar in both
bands, whereas the amplitude of the spectral variations is much
smaller. We conclude
that the  flux variations in \pks\ are  associated with spectral variations
which are significant, although of smaller amplitude. 

\section{Spectral variability: model-independent analysis}

Previous X-ray studies, based mainly on time-averaged spectral studies, have
offered different interpretations of the physical origin of the X-rays in \pks.
Our main aim in this work is to investigate whether the study of the source's
X-ray spectral variability can help us clarify the situation. 

We begin our study with the use of simple methods and tools
like flux-flux plots, hardness ratio versus count rate plots, and 
the fractional variability versus energy plots. These can provide useful
information without  any a priori assumption regarding the shape of the X-ray
continuum spectrum. Thus, the results from the study of these plots
can be considered as ``model-independent". 

In the following analyses we fit various data sets with linear functions, using
the the routine \verb+fitexy+ (Press et al. 1997) that accounts for the errors
not only on the y-axis but along the x-axis as well. In many plots, we observe
that the data follow a linear trend, but with substantial scatter around it. In
this case, in order to reveal clearly the underlying trend, we bin the data using
count rate bins  of fixed size (0.1 counts/s), and
compute the weighted mean (along both the y and x-axis) of all the points
which fall into a bin. The error on the weighted mean is computed
following Bevington's prescriptions (Bevington 1969).

\subsection{The Flux-Flux Plot}

This simple method of plotting the hard versus the soft 
X-ray count rate can be used to investigate the presence of spectral variations
(Churazov et al. 2001; Taylor et al. 2003).
For example, if the flux-flux plot of an object is well described by a simple
linear model (i.e. a straight line) then the continuum varies only in
normalization and not in shape. Furthermore, if the intercept of the y--axis is
non-zero, then there also exists a second spectral component, which is non
variable, neither in shape nor in normalization. On the other hand,  if the
flux-flux plot  is best described by a non-linear function (a power law for
example), then the observed flux variations should be associated with spectral
variations as well, i.e. we may be dealing with a power-law like continuum with
variable normalization and photon index as well. 

We made a flux-flux plot for \pks\ (see Fig.~\ref{figure:fig3}) using the
2.5--5 keV and 5--15 keV light curves as representative of the ``soft" and
``hard" energy band flux, respectively. The gray (light blue in color) symbols
represent the original data with time bins of two days.

A visual inspection of Fig.~\ref{figure:fig3} clearly shows
that there exists a linear relationship between the soft and hard flux (this is
true also for the separate S1 and S2 data sets). However, the fit of the 
simple linear model is poor, due to the scatter in the hard flux values for a
given soft flux count rate.  Nevertheless, the scatter is small and contains no
systematic deviations from a linear model. To verify this, we have also plotted
the binned data (black filled circles in Fig.~\ref{figure:fig3})  together with the best-fit
linear models.
The linear model fits the binned data very well, and the best fitting results for
binned data sets are consistent with each other. In all cases (S1, S2, and 
combined data sets) the y--axis intercept is consistent with zero. Similar results are
obtained when we use other bands (either than $5-15$ keV) as representative
of the high energy band. In all cases, the soft vs. hard band plots are well
fitted, on average, by straight lines  with almost zero y--axis intercepts. 
The best fitting results are listed in Table 1. 

In conclusion, we find that the hard and soft band light curves are very well 
correlated. On average, this correlation is well described by a simple linear
relation with intercept around zero. This relation is the same for both the S1
and S2 data sets, which implies that the source behaves in the same way during
the long (weeks, years) and shorter (days) time scales.  The linear hard vs
soft relation  is consistent with a power-law like X-ray continuum which is
variable only in normalization and not in shape. The zero y--axis intercept
also suggests that there is no indication of an extra, constant component (like
e.g. the constant disk-reflection component that has been detected in the X-ray
spectra of a few nearby AGN). When we examine the full data sets, we find
moderate scatter of the points around the underlying, linear, hard vs soft
relation. This result implies some weak spectral variability which is
uncorrelated with flux.

\subsection{Hardness Ratio vs. Count Rate}

Figure~\ref{figure:fig4} shows the Hard/Soft X-ray color plotted versus the
count rate for unbinned (gray/blue crosses) and binned data (black symbols), 
respectively. 
A visual inspection of Figure~\ref{figure:fig4} clearly shows
that the hardness ratio does not appear to correlate with the count rate
(S2 and S1 data sets are virtually indistinguishable in this analysis). 
However, when we
fit a simple linear model to the data, we do  observe a weak positive
correlation (see Table 1).  The best fitting slope is
$0.05\pm0.02$, which is a $2.5\sigma$ effect. 

Better insight in the presence of correlations between $HR$ and total count
rate can be obtained by investigating the binned data  (black circles in 
Fig.~\ref{figure:fig4}).  
In this case, the best fitting  is $0.06\pm 0.02$, which apparently confirms the presence
of a positive correlation at  the 3$\sigma$ level. 
However, we note that the linear best-fit results are strongly dependent on the
data points with  count rate below $\sim 1-1.2$ count/s. These bins are
scarcely populated. Indeed, if we take into account only the  binned points
that contain at least 8 data, the significance of the
positive correlation disappears (see Table 1) and the HR values are not
correlated with the count rate any more. This result is consistent with the
findings from the previous section. 

In summary, the results from the study of the  $HR-ct$ plots confirm the
findings from the study of the  flux-flux plots. The spectral behavior
of the source  is identical on different time scales.  We find no significant
evidence for positive or negative correlations between hardness ratio and
total count rate. As before, this result suggests that the flux variations
occur while the  continuum spectral shape remains constant. 

We do note though that there is a  hint of a weak,  positive trend between the
hardness ratios and flux, in the sense that  the spectrum hardens as the
source brightens. This trend is more obvious at low flux levels. We need to
accumulate more data points at high and (mainly) low count rates, in order to
be able to confirm (or refute) this hypothesis on statistical grounds.

\subsection{Fractional Variability vs. Energy}

Another simple way to quantify the variability properties of \pks\, without
considering the time ordering of the values in the light curves, is based on
the fractional variability parameter $F_{\rm var}$ (e.g. Rodriguez-Pascual et
al. 1997; Vaughan et al. 2003).  This is a commonly used measure of the intrinsic
variability amplitude relative to the mean count rate, corrected for the effect
of random errors, i.e., \begin{equation} F_{\rm
var}={(\sigma^2-\Delta^2)^{1/2}\over\langle r\rangle} \end{equation} 
\noindent
where $\sigma^2$ is the variance, $\langle r\rangle$ the unweighted mean count
rate, and $\Delta^2$ the mean square value of the uncertainties associated with
each individual count rate. The error on $F_{\rm var}$ has been estimated following
Vaughan et al. (2003):
\begin{equation}
\sigma_{F_{\rm var}}= \sqrt{\left( \sqrt{\frac{1}{2N}}\cdot\frac{\Delta^2}{\langle r\rangle^2 F_{\rm var}} \right)^2+\left( \sqrt{\frac{\Delta^2}{N}}\cdot\frac{1}{\langle r\rangle}   \right)^2}
\end{equation}
\noindent
where $N$ is the number of data points.

We computed $F_{\rm var}$ on selected energy bands.  These bands
were chosen in such a way to have mean count rates similar and sufficiently
high. 
A separate analysis for S1 and S2 suggests a similar behavior for the two data sets.
This conclusion is confirmed in a
more quantitative way by computing the difference  $\sum_{i} \vert F_{\rm
varS1} (E_{\rm i}) - F_{\rm varS2} (E_{\rm i})\vert$ over all  energies, which
yielded $0.43\pm0.17$, which is a 2.5$\sigma$ effect.

Having established that $F_{\rm var}$(S1) $\sim$ $F_{\rm var}$(S2), we can
estimate the $F_{\rm var}$ at each energy for the combined data set, using time bins
of two days. The results,
plotted in Fig.~\ref{figure:fig5}, suggest that in \pks, 
the fractional variability amplitude is  nearly constant between 2 and 15 keV.
These findings are consistent
with the results from the previous sections in that they indicate that there exist
no flux-related, large amplitude spectral variations.

\section{Time-resolved Spectral Analysis}

Since the data consist of short snapshots spanning a long temporal baseline,
they are well suited for monitoring the spectral variability of \pks\ but, in
principle, not for a time-averaged spectral analysis. Nevertheless,  since the
spectral variability is not very pronounced (see $\S4$), we have carried out a
time-resolved spectroscopy by averaging spectra over two-month intervals. This
choice is a trade-off between the necessity of  accumulate sufficient counts
for a reliable spectral analysis and the need to use limited temporal intervals
to minimize the effects of the slow drift in the PCA gain. 

We fitted all six two-month spectra with a simple power-law (PL) model 
absorbed by Galactic $N_{\rm H}$ ($4.8\times10^{20}{\rm~cm^{-2}}$). The model fits all
the data reasonably well. The best-fit results are listed in Table 2 and, as an
example, Fig.~\ref{figure:fig6} shows the best-fit PL model (together with the
best-fit residuals) to the first 2-month long spectrum of the source. 

Assuming $H_0=71{\rm~km~s^{-1}~Mpc^{-1}}$, $\Omega_\Lambda=0.73$ and
$\Omega_{\rm M}=0.27$ (Bennet et al. 2003), the intrinsic luminosity is L$_{\rm
2-10~keV}\sim 9 \times 10^{44}$ \lum\ (the corresponding flux is reported
in Table 2), which corresponds to a sizable fraction
($>15$\%) of the Eddington luminosity and causes the bolometric luminosity to
exceed significantly $L_{\rm Edd}$ if we assume that $L_{\rm 2-10~keV}\sim
0.05-0.10~L_{\rm bol}$.

The best-fit spectral slope values are rather steep, as expected for a NLS1
object. They are all consistent within the errors, suggesting that there are no
spectral variations, in agreement with the results in  $\S4$. Furthermore, we
find no evidence for a second spectral component nor for the presence of an Fe
K$\alpha$ line  in any of the spectra.  Adding a Gaussian line at 6.4 keV to
the PL continuum model  does not improve the fit significantly in any of the
six spectra. 

In order to investigate the issue of spectral variability in \pks\ in greater
detail, we also extracted energy spectra from each individual PCU2 pointing. 
All spectra are reasonably well fitted by a simple power law absorbed  by
Galactic $N_{\rm H}$. 
In Figure~\ref{figure:fig7} we plot the best-fit  $\Gamma$ versus time
for both the individual pointings (gray crosses) and the 2-month averaged spectra 
(black filled circles). Clearly, this
plot suggests that the continuum spectral shape does not change with time. We
estimate a $\chi^2=164.7$ for 207 dof and a $\chi^2=2.7$ for 5 dof in the case
of the individual and the 2-month averaged $\Gamma$ light
curves, respectively. The weighted averages for the photon index
are $\langle\Gamma\rangle=2.20\pm0.03$ for the 2-month intervals and 
$\langle\Gamma\rangle=2.23\pm0.02$ for the individual spectra.

Despite the absence of substantial spectral variability, a time-averaged 
analysis of the whole monitoring campaign is not feasible, because the continuous drift 
in the detector gain hampers the creation of an appropriate response matrix.
Nevertheless, we have tried to exploit spectrally all the available data by fitting simultaneously 
the six 2-month averaged spectra. The resulting photon index is $\Gamma=2.20\pm0.3$, 
fully consistent with the weighted mean. However, no tighter
constraints on the presence of the Fe K line are obtained with this analysis.

In conclusion, we find that the spectrum of \pks\ is well fitted by a simple PL
model with steep photon index, and shows no spectral slope variations.

\section{Discussion} 

\begin{figure}
\begin{center}
\includegraphics[bb=105 25 425 455,clip=,angle=0,width=12.cm]{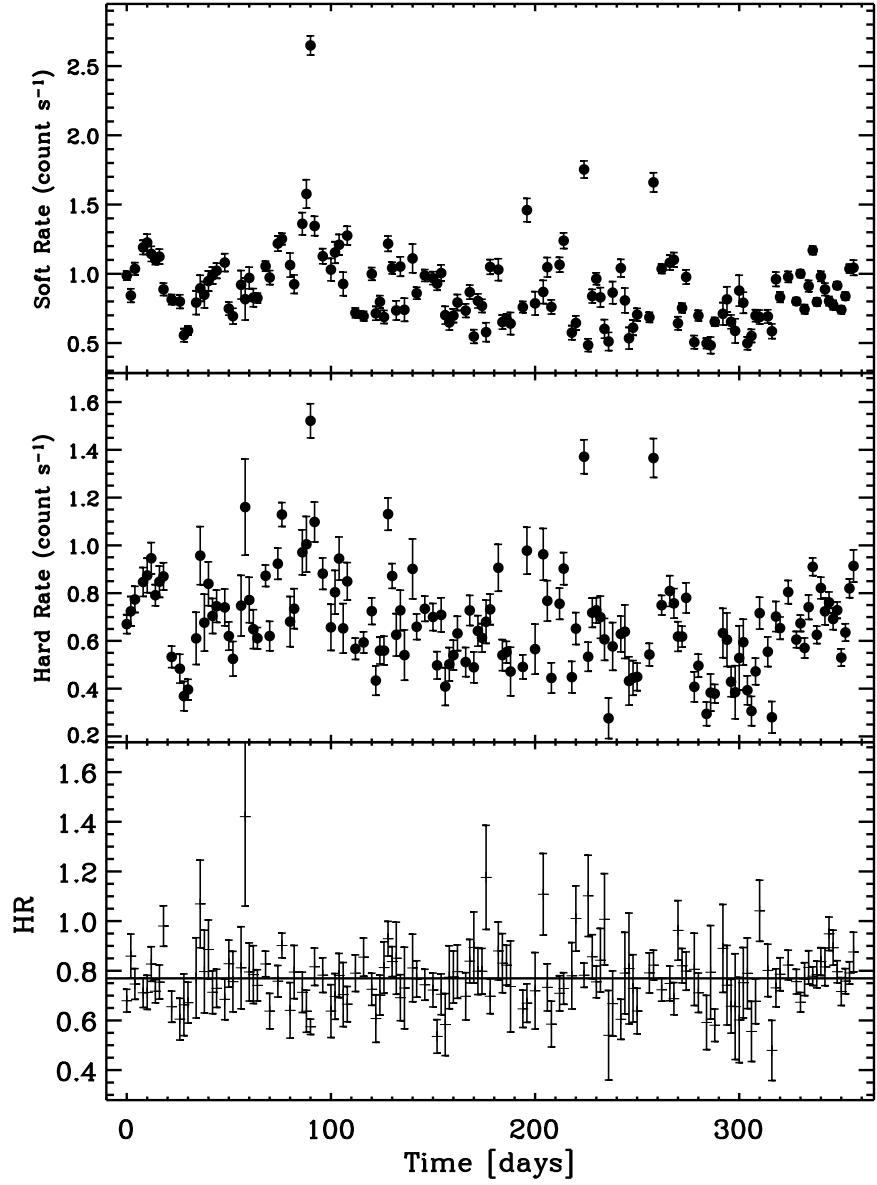}
\caption{Top panel: Soft (2.5--5 keV) light curve of \pks\ using PCU2 data only.
Middle panel: Hard (5--15 keV) light curve.
Bottom panel: Hardness Hard/Soft ratio light curve; the solid line represents the 
average value of the hardness ratio. Time bins are two days.} 
\label{figure:fig2}
\end{center}
\end{figure}

\begin{figure}
\begin{center}
\includegraphics[bb=45 32 350 300,clip=,angle=0,width=9.cm]{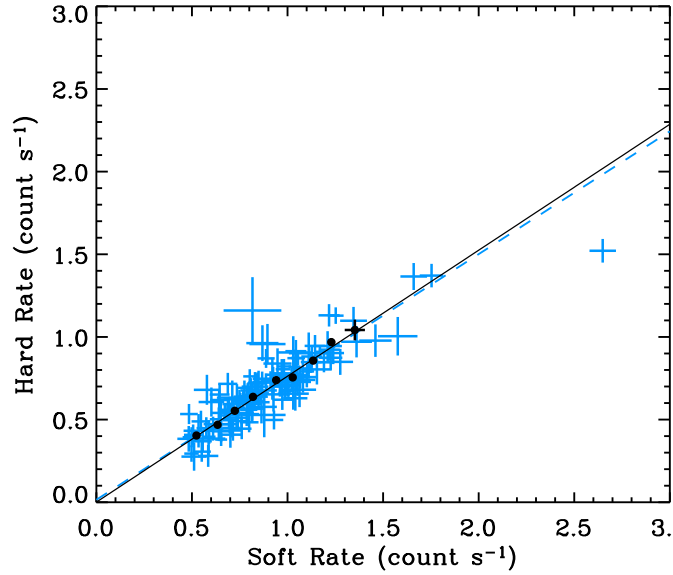}
\caption{Flux-flux plot of \pks. The soft count rate refers to the 2.5--5 keV band,
the hard to the 5--15 keV band. The gray (blue in color) symbols correspond 
unbinned data points, whereas the black filled circles refer to the weighted
mean of the original points with fixed bins of 0.1 c/s. The outlier corresponds to
the point with the highest count rate occurring $\sim$ 90 days after the beginning
of the monitoring campaign. The gray dashed line represents the best-fit linear
model for the unbinned points, whereas the black continuous line refers to the
binned data points.} 
\label{figure:fig3}
\end{center}
\end{figure}

\begin{figure}
\begin{center}
\includegraphics[bb=45 32 350 300,clip=,angle=0,width=9.cm]{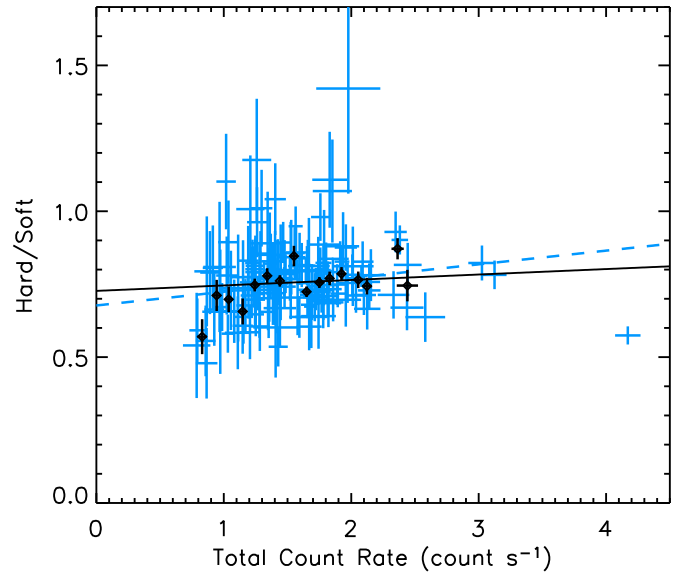}
\caption{Hardness ratio (5--15 keV/2.5--5 keV) plotted versus the count rate.
The gray (blue in color) crosses refer to unbinned data, whereas the black 
filled circles correspond to weighted means of the original points with 
fixed bins of 0.1 c/s. The gray dashed line represents the best-fit linear
model for the unbinned points, whereas the black continuous line refers to the
(well-populated) binned data points; see text for more details.}
\label{figure:fig4}
\end{center}
\end{figure}

\begin{figure}
\centering
\includegraphics[bb=45 35 350 290,clip=,angle=0,width=9.cm]{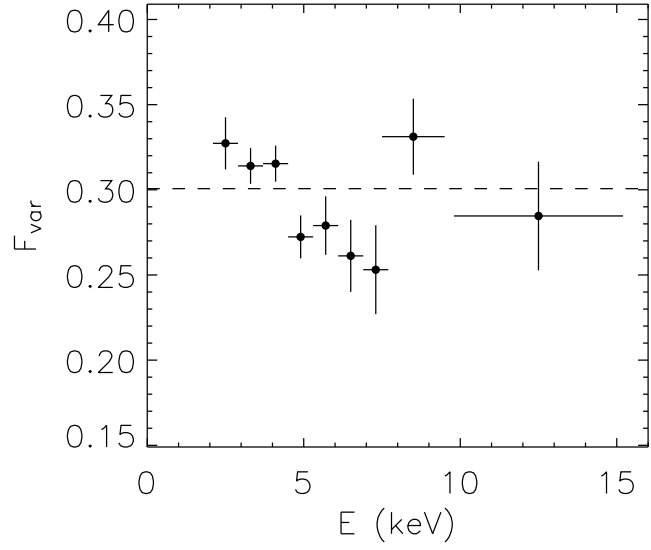}

\caption{Fractional variability amplitude as a function of
energy for \pks. 
The error-bars along the x axis simply represent
the energy band width. The error bars along the y axis are computed following
Vaughan et al. 2003. The dashed line represents the weighted mean of $F_{\rm var}$.
} 
\label{figure:fig5}
\end{figure}

\begin{figure}
\includegraphics[bb=65 5 575 715,clip=,angle=-90,width=16.cm]{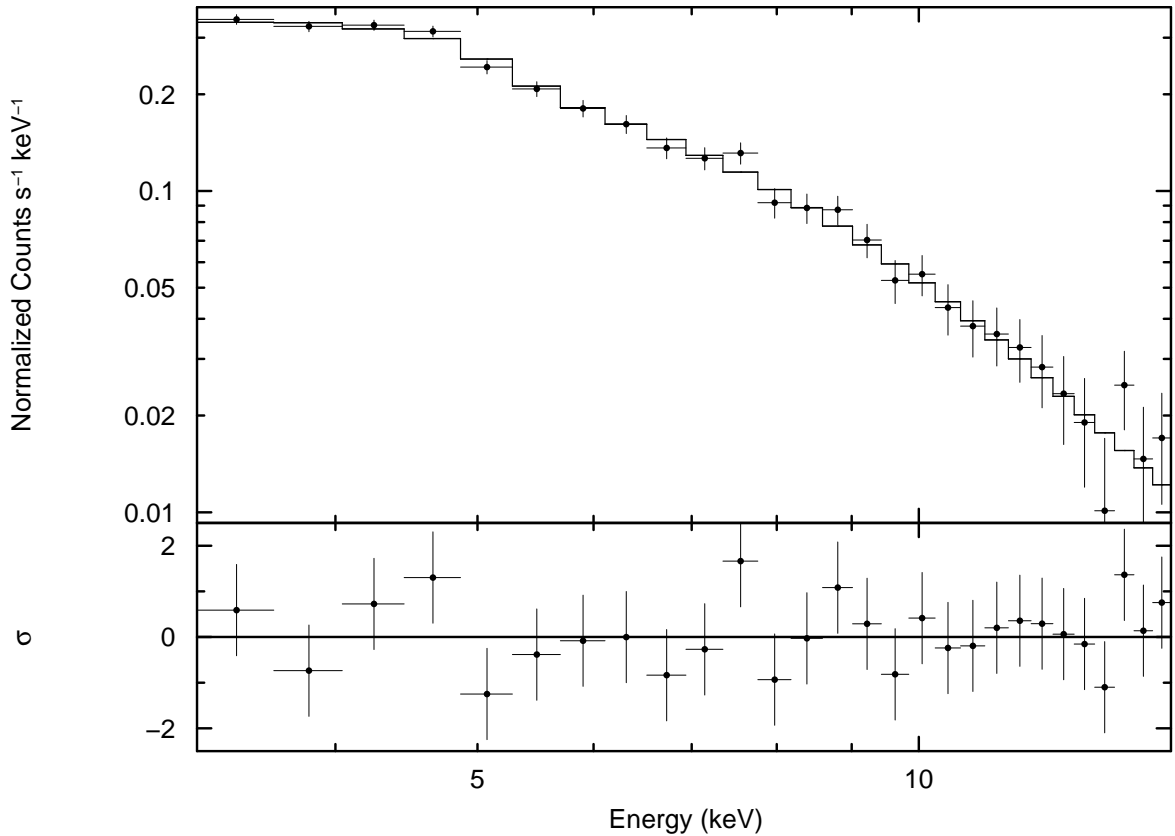}
\caption{PCA spectrum of \pks\ during the first 2-month interval obtained using PCU2 data only} 
\label{figure:fig6}
\end{figure}

\begin{figure}
\begin{center}
\includegraphics[bb=75 60 460 370,clip=,angle=0,width=12cm]{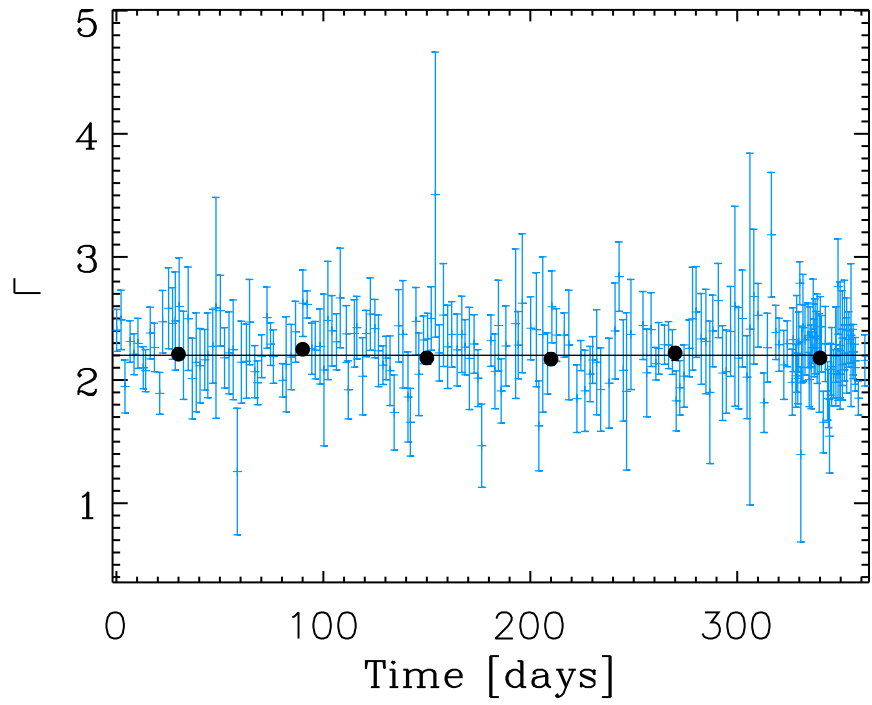}
\end{center}
\caption{Gray (blue in color) symbols represent the values of $\Gamma$ 
obtained in individual pointings plotted against time. Black filled circles refers to 
2-month averaged spectra. All error-bars are 1$\sigma$. The solid line represent the 
mean value obtained averaging the 2-month spectra.}
\label{figure:fig7}
\end{figure}

We have used data from a year-long \rxte\ monitoring campaign to study the 
spectral variability of \pks\, following model-independent and spectral model 
fitting methods. The main results can be summarized as follows:

\begin{itemize}

\item The light curves (Fig.~\ref{figure:fig2}) show large amplitude variations
on time scales ranging between half a day to several days/weeks,  independently
of the sampling pattern and energy range. Indeed, the fractional variability
$F_{\rm var}$ is of the order of 25-30\% in all energy bands (Fig.~\ref{figure:fig5}).

\item  The large amplitude flux variations are not accompanied with substantial
spectral variations. We do observe some low amplitude spectral variations, but they are
not flux correlated (as both the flux vs flux and $HR-ct$ plots show).
There is a hint of a positive trend in the
$HR-ct$ plot, mainly at low fluxes (Fig.~\ref{figure:fig4}) but more data are
needed to confirm this result.

\item The spectra, both from individual orbits and from the 2-month averaged
data,  are well fitted by a power-law model, with $\langle\Gamma\rangle\simeq2.2$.
This average value is the same for both S1 (long time scales) and S2 (short time scales) data sets.

\item We find no indication of an iron line, in any of the 2-month averaged or
individual spectra. The weakness of the reflection features in the
2--15 keV range  is confirmed by the flux-flux plot (Fig.~\ref{figure:fig3}).
The almost zero intercept in the flux-flux plot suggests a negligible contribution 
from a constant reflection component which is usually observed in other Seyfert
galaxies.
\end{itemize}

The time-averaged spectral results of \pks\ ($\Gamma= 2.2$, EW $<$ 50 eV), obtained in 
the 2.5--15 keV energy range during the \rxte\ monitoring campaign, 
are in broad agreement with the findings obtained by previous X-ray missions. 
A compilation of the measured photon indices 
of \pks\ over one decade is
reported in Gliozzi et al. (2000): with the exception of \rosat\ ($\Gamma\simeq 3.1$;
Brinkmann et al. 1997) that only probes the soft range 0.1-2.4 keV, all the values 
obtained by ${\it GINGA}$ ($\Gamma\simeq 2.24$; Remillard et al. 1991),
${\it EXOSAT}$ ($\Gamma\simeq 2.21$; Lawson et al. 1992), and \asca\
($\Gamma\simeq 2.25$; Leighly et al. 1999) are consistent with the \rxte\ findings,
suggesting that the spectrum of \pks\ has remained roughly constant for more than 15 
years. The average spectral slope measured with \rxte\
is also in good agreement with the best-quality spectral data to date obtained with \xmm\
in the 2--10 keV range ($2.12\pm0.15$; O'Brien et al. 2001).
In addition to \pks, Leighly (1999) presents a comprehensive analysis of 22
NLS1 observed by \asca, showing that the photon index of \pks\ is fully consistent
with the average $\Gamma$ of the NLS1 sample. On the other hand, the equivalent
width of an Fe K line of \pks\ measured by \asca\ (EW= 106  eV) has the lowest value 
among the NLS1 with detected lines (but see Ballantyne et al. 2002 for a different 
interpretation of the \asca\ data). The weakness of the Fe K line
is also consistent with the low upper limits obtained by \xmm:
$<$ 10 eV for a narrow line ($\sigma=10$ eV) and $<$ 90 eV for a broad line 
($\sigma=300$ eV; O'Brien et al. 2001). 

The \rxte\ spectral variability findings (large flux changes without substantial
spectral variations) can be compared with the work of Brinkmann et al. (2004),
who present the results from several \xmm\ observations of \pks\ carried out
over a period of almost two years. Despite the fact that the \xmm\ observations are
carried out in a different energy band (0.2--10 keV versus 2.5--15 keV in our case) and probe
shorter time scales, we find a remarkable agreement with the \rxte\ results. In particular,
Figure 4 of Brinkmann et al. (2004) shows that large amplitude count rate changes are not 
accompanied by a clear spectral trend, and that there is a substantial scatter in the
hardness ratio values for a given count rate.

Since the primary goal of this work was to investigate the origin of the X-rays
in \pks\ and in particular to assess the role played by the putative jet, it
is  instructive to compare our study with similar analyses carried out both on
radio-quiet  AGN (where the X-rays are thought to be produced by Comptonization
in the corona that is closely connected with the accretion disk) and on
radio-loud jet-dominated AGN (whose radiation over the entire energy range is
ascribed to jet emission).

Previous studies of nearby radio-quiet AGN (including mostly broad-line Seyfert
galaxies, but also NLS1), based on yearly-long
\rxte\ observations, have revealed a well defined spectral behavior, which is
characterized by: 1) A nearly ubiquitous spectral softening when the source
brightens (Papadakis et al. 2002; Markowitz \& Edelson 2004). 2) A more
pronounced variability in the soft (2--4 keV) band than in the hard (7--10 keV)
energy range (Markowitz \& Edelson 2001). 3) Time-resolved spectra with
prominent reprocessing features: an Fe K$\alpha$ line is always statistically
required by the fit and the reflection fraction $R$ is typically of the order
of 0.5 (Markowitz et al. 2003). 

Also radio-loud jet-dominated AGN, such as the brightest blazars, have been
targeted by several monitoring campaigns with \rxte\ in recent years. However,
most of these studies are focused either on the interpretation of the
broad-band spectral energy distribution  
or on the variability
properties with structure functions and power  density spectra (e.g., Kataoka
et al. 2001). Recently, a model-independent study of the  spectral variability
properties of the nearby blazar Mrk~501 has been performed by  Gliozzi et al.
(2006a). Interestingly enough, Mrk~501 shows a markedly different spectral
behavior compared to radio-quiet objects, with spectral variations
characterized by a hardening when the source brightens, and with the fractional
variability more pronounced at higher energies. The bluer when brighter behavior
has been observed also in other blazars (see Rebillot et al. 2006 and references therein).

If we compare the spectral properties of \pks\ with those of radio-quiet AGN and
blazar objects, we find that they are not consistent with the characteristic
behavior of either AGN class. Despite the significant flux variations, \pks\
does not show any clear softening (or hardening) of the spectra when the count
rate increases, nor does it present any marked energy dependence of $F_{\rm var}$. 
This suggests that \pks\ is a remarkably peculiar object.

Another object that shows a similar spectral behavior is 3C~273.
Long-term \rxte\ observations reveal a 
dual spectral behavior for 3C~273, with achromatic flux variations (which have been
interpreted as associated with
jet emission) during 1996--1997 and Seyfert-like spectral variations (softening of
the spectrum when the source brightens) accompanied by Fe K$\alpha$ detections
during 1999-2000 (Kataoka et al. 2002). This suggests, according to Kataoka et al. (2002),
that 3C~273
is a hybrid object with contributions from the jet and the accretion-related component
varying with time. However, it is not just 3C 273 that
shows a similar achromatic variability behavior. For example, the
radio-quiet quasar PG~0804+761 shows flux variations without any spectral variation associated
and has an iron line at $\sim$6.4 keV with EW$\sim$110 eV in its spectrum (Papadakis et al. 2003).

The presence of a jet in \pks, probably seen at a small viewing angle, is
supported by the flat radio spectrum and the limited radio extension  of the
source  at VLA resolution (Lucyna Kedziora-Chudczer private communication).  In
addition, the lack of reprocessing features in the \rxte\ PCA spectrum
(the Fe K line is never detected at significant level and EW\ltsima 50 eV at a 90\% confidence
level) also favors the presence of a jet, which may
actually be the dominant source of X-ray emission. This conclusion can find further support
from the similarity of \pks\ with 3C~373 during its supposedly jet-dominated period that is characterized by
substantial flux changes without spectral variability. Note that, contrary to 
\pks, 3C~373 sometimes shows a fairly strong Fe K line in 
its spectrum (EW$\simeq$ 100--200 eV) and Seyfert-like spectral variations, suggesting that
\pks\ might be more jet-dominated than 3C~273. 

In the framework of jet-dominated models, flux
variations without relevant spectral changes can be naturally explained by changes in the 
electron injection rate (Kirk, Rieger, \& Mastichiadis 1998) or by changes in the orientation
of the jet (e.g., Villata \& Raiteri 1999).

Alternatively, in the framework of (thermal) coronal models, the spectral variability is essentially 
regulated by two parameters: the temperature $kT$ and the optical depth $\tau$. In principle, one might
reproduce the peculiar spectral variability shown by \pks\ by  assuming that the
respective changes in  $kT$ and $\tau$ cancel out. However, considering the large
amplitude flux variations observed in \pks, we regard this hypothesis as unlikely,
since it requires fine tuning. 

Nonetheless, coronal models may still represent a
viable solution for \pks, if we allow for changes in the supply of seed photons and variable
power in the Comptonized component (see Gierli\'nski \& Zdziarski 2005 and references
therein). Indeed, spectral behaviors quite similar to the one shown by \pks\ (in particular,
$F_{\rm var}(E)=$constant) have been observed in Galactic black hole systems
(GBHs) during their transition from the hard to the soft spectral state, and
interpreted in the framework of coronal models (Gierli\'nski \& Zdziarski 2005).
Interestingly, in this scenario, \pks\ would be associated with GBHs in their 
intermediate states, where sometimes transient relativistic jets are observed
(Fender, Belloni, \& Gallo 2004). In the coronal scenario, the complete lack of reprocessing 
features in the energy spectra of \pks\ might be explained by extreme relativistic blurring
(Crummy et al. 2006) or by very strong Comptonization (Matt et al. 1997).

In summary, \pks\ shows unique X-ray properties, which are not similar to
what most (but not all) of radio-quiet Seyferts and blazars show. It
looks similar to 3C~273 (during the supposedly jet-dominated period), 
in the sense that this source also does not show
spectral variations, but the average $\Gamma$ in 3C~273 is much flatter than
in \pks. It also looks similar to the radio-quiet quasar PG~0804+761, and in this case, 
also the spectral slopes are similar. However, contrary to \pks, PG~0804+761 shows a standard
Fe K$\alpha$ line. If jet
emission is dominant, then the jet varies in a non-standard way (as
otherwise, \pks\ should show the bluer when brighter behavior). If the corona
is dominant, then again, the source must be in this ``rare" intermediate 
state, as it is different to most of the other Seyfert galaxies (both broad and
 narrow-lined). 
In either case, \pks\ is a rare source, a very interesting one. With
the new data from the current observations, we will: a) check that the first year results are
indeed representative of the source's behavior, b) we will study
better the possibility of weak flux related spectral variations, and c)
we will study the flux variations in terms of power spectrum analysis.
This will help us decide which mechanism is dominant in the X-rays.

\begin{acknowledgements} 
MG acknowledges support by the RXTE Guest Investigator Program
under NASA grant 200857. IEP and WP gratefully acknowledge travel support through the bilateral
Greek-German IKYDA project based personnel exchange program. We thank the referee for his
comments and suggestions that improved the clarity of the paper.
\end{acknowledgements}

\begin{table} 
\caption{Model-independent Spectral Variability}
\begin{center}
\begin{tabular}{lcccc}
\multicolumn{5}{c}{{\bf Flux -- Flux}}\\ 
\hline        
\hline
\noalign{\smallskip}        
Data            &          Slope  &  Intercept    &  $\chi^2$/dof &     $P_{\rm\chi^2}$            \\
\noalign{\smallskip}
\hline  
\noalign{\smallskip}
unbinned      & 0.74$\pm$0.03  &  0.02$\pm$0.02   & 218.7/139  &  $1.9\times 10^{-5}$   \\
\noalign{\smallskip}
binned        & 0.76$\pm$0.03  &  0.00$\pm$0.03   & 8.7/7    &  $2.8\times 10^{-1}$   \\
\noalign{\smallskip}
\hline
\noalign{\smallskip}
\noalign{\smallskip}
\multicolumn{5}{c}{{\bf HR -- rate}}\\ 
\hline        
\hline
\noalign{\smallskip}        
unbinned      & 0.05$\pm$0.02    &   0.68$\pm$0.03   & 172.6/138  &  $2.4\times 10^{-2}$   \\
\noalign{\smallskip}
binned        & 0.06$\pm$0.02    &   0.65$\pm$0.03   & 30.2/14  &  $7.1\times 10^{-3}$   \\
\noalign{\smallskip}
binned$^\clubsuit$   & 0.02$\pm$0.03   &   0.73$\pm$0.06     & 5.8/5  &  $3.2\times 10^{-1}$   \\
\noalign{\smallskip}
\hline
\end{tabular}
\end{center}
\footnotesize
$^\clubsuit$ Only bins well populated ($>8$ data points)
are considered
\label{tab1}
\end{table}

\begin{table} 
\caption{Spectral analysis results}
\begin{center}
\begin{tabular}{ccccc} 
\hline        
\hline
\noalign{\smallskip}        
Interval &  $\chi^2$/dof & $F_{\rm 2-10~keV}$                 & $\Gamma$                 & EW$_{\rm FeK\alpha}$ \\
\noalign{\smallskip}
(yy/mm/dd)         &                        & $(10^{-11}{\rm erg~cm^{-2}~s^{-1}})$ &                        &  (eV)   \\
\noalign{\smallskip}
\hline  
\noalign{\smallskip}
05/03/04-05/05/03       & 15.6/25               &  1.88                                & $2.21_{-0.14}^{+0.09}$ &  $< 155$   \\
\noalign{\smallskip}
\hline
05/05/05-05/07/04       & 18.6/25               &  2.20                                & $2.25_{-0.07}^{+0.07}$ &  $< 68$   \\
\noalign{\smallskip}
\hline
05/07/06-05/09/04       & 32.1/25               &  1.77                                & $2.18_{-0.07}^{+0.08}$ &  $< 35$   \\
\noalign{\smallskip}
\hline
05/09/06-05/11/05       & 22.9/25               &  1.77                                & $2.17_{-0.11}^{+0.11}$ &  $< 52$   \\
\noalign{\smallskip}
\hline
05/11/07-06/01/06       & 31.8/25              &  1.54                                & $2.22_{-0.10}^{+0.08}$ &  $< 48$   \\
\noalign{\smallskip}
\hline
06/01/08-06/03/02       & 31.7/25              &  1.88                                & $2.18_{-0.05}^{+0.05}$ &  $< 50$   \\
\noalign{\smallskip}
\hline

\end{tabular}
\end{center}
\label{tab1}
\end{table}       

\end{document}